# Helicity of mean and turbulent flow with coherent structures in Rayleigh-Bénard convective cell

A. Eidelman[1, a], T. Elperin[1,b], I. Gluzman,[1,c] and E. Golbraikh[2, d]

[1]Mechanical Engineering Department and  [2]Department of Physics,

Ben-Gurion University of the Negev, Beer-Sheva, 84105, Israel

We present results of the study of a turbulent air flow with a large scale circulation in Rayleigh-Bénard rectangular convective cell with a heated bottom wall and a cooled top wall. Velocity fields were measured using Particle Image Velocimetry in two sets of mutually perpendicular planes parallel to the vertical walls of the cell. Experiments revealed the existence of the main roll, having a length scale of the order of the size of the cell, and elongated eddy rings adjacent to the bottom and top of the main roll. The mean horizontal velocity of the main roll and the mean vorticity of eddy rings are almost aligned in a large part of the flow. The helicity of the mean flow is quite high, and is the source of turbulent helicity. Since helicity of the mean flow and turbulence is quite large, the flow in Rayleigh-Bénard convective cell is well suited to study properties of helical turbulence. Spatial distribution of the turbulent kinetic energy is almost locally isotropic in the central region of the cell. Spectra and cross spectra of turbulent velocities reveal two distinct ranges in the inertial interval with the slopes close to $-5/3$ and $-7/3$. We believe that emergence of these two intervals is associated with energy and helicity cascades that affect turbulence. We determined also turbulent helicity using the measured velocity cross-spectra. We found that the magnitude of the length scale where the slope of the velocity spectra changes and the magnitude of the length scale defined as the ratio of turbulent energy to the helicity are approximately the same. The slopes of power law spectra of helicity in the intervals above and below the transition length scale are equal to $-2/3$ and $-4/3$, respectively. Remarkably, similar inertial sub-ranges in turbulent energy spectra were observed in various laboratory and geophysical turbulent flows.

---

[a] Electronic mail: eidel@bgu.ac.il  
[b] Electronic mail: elperin@bgu.ac.il URL: http://www.bgu.ac.il/me/staff/tov  
[c] Electronic mail: igal.gluzman@gmail.com  
[d] Electronic mail: golbref@bgu.ac.il

I. INTRODUCTION

Helicity of a flow is defined as $H_e = \vec{u} \cdot \vec{\omega}$ where $\vec{u}$ is a velocity and $\vec{\omega}$ is a vorticity of the flow. Mean turbulent helicity $H_t = \frac{1}{2}\langle \vec{u}'(\vec{r},t) \cdot \vec{\nabla} \times \vec{u}'(\vec{r},t)\rangle$, where $\langle...\rangle$ denotes averaging over the statistical ensemble or volume, $\vec{u}'(\vec{r},t)$ is turbulent component of the velocity, is an invariant of the Euler equations along with energy[1]. Investigations of helicity of turbulent flows were motivated first by their ability to generate secondary magnetic fields in conducting media[2]. Studies of the role of helicity in turbulence were stimulated by the paper by Moffatt [3] who showed that helicity characterized the degree of linkages of vortex lines which move with the fluid, and established helicity invariance for barotropic inviscid flow. Since then helicity has the status comparable with that of classical invariants as energy, angular momentum and others.

Based on an analysis of two helical waves, Kraichnan [4] suggested that energy transfer should be smaller in turbulence with the maximum helicity for a given energy spectrum shape. This suggestion was supported by Andre and Lesieur [5] numerical simulations of three-dimensional isotropic turbulence at very high Reynolds number. Clearly analysis of helical turbulence requires investigating fundamental properties of turbulent flows. Tsinober and Levich [6] conjectured that helicity in turbulent and transitional flows with three-dimensional coherent structures is of special importance. Nevertheless, the role of helicity in turbulent flows is still a subject of vigorous discussions. Some researchers believe that helicity has dynamical significance in turbulent flows, while helicity was shown to be dynamically irrelevant in some cases (see e.g. review in Galanti and Tsinober [7]). Moiseev et al. [8] and Lilly [9] hypothesized that helicity is probably involved in generation of hurricanes in the atmosphere. The latter conjecture is related with inhibition of turbulent energy cascade or even reversal of the direction of energy cascade in helical turbulence (see e.g. Belian et al. [10] and references therein).

Energy spectra of turbulence with the nonzero mean turbulent helicity were first studied by Brissaud et al. [11], where helicity dissipation rate, $\eta$, together with turbulent energy dissipation rate, $\varepsilon$, were used as parameters. This study revealed the existence of turbulent energy spectrum that is affected by helicity cascade, and has a slope $-7/3$ ($E(k) \sim \eta^{2/3} k^{-7/3}$) that appears along with the Kolmogorov's slope $-5/3$. The spectrum having the slope −7/3 characterizes the absence of energy cascade ($\varepsilon = 0$) according to the approach based on the effective correlation time.

Moffatt [12] indicated the need to determine the helicity spectrum and mentioned two methods for helicity measurements – direct measurements of helicity H based on its definition:

$$H = \langle \vec{u}'(\vec{x}) \cdot \vec{\omega}(\vec{x} + \vec{r}) \rangle \qquad (1)$$

or using cross-correlation function of velocity.

Helicity was directly measured only in a few studies which employed complicated probes for determining velocity and its spatial derivatives. Two types of probes were designed for such measurements. Wallace [13] employed an array of nine thermo-anemometer probes with wires directed under different angles with respect to the mean velocity. Tsinober et al. [14] combined measurements of the longitudinal component of the velocity with a hot-film sensor and determined vorticity using seven-point potential measurements of the electric field in salt water flowing in a magnetic field. Kholmyansky et al. [15] measured helicity in a turbulent grid flow in the salt water tunnel with the potential probe having the size of 2 mm. Experiments were conducted downstream of grids with square and circular meshes. The sign of the mean helicity was determined by controlling direction of rotation of propellers installed in grid holes. In all cases the main contribution to the mean helicity came from the large scales of the order of the diameter of the tunnel, and the direction of propellers rotation did not affect the sign of the mean helicity. This observation was indication that small disturbances of helicity were amplified by the turbulent flow past the grid. Wallace et al. [16] examined properties of the normalized helicity density in turbulent boundary layer, in two-stream mixing layer, and in grid flows measured with nine-sensor hot-wire probe. The conclusion was that the PDF of the normalized helicity density in turbulent shear flows showed a tendency for the alignment of the velocity and vorticity vectors, although this tendency was less pronounced for the fluctuating components. Kholmyansky et al.[17] reported results of helicity measurements in jet and wake flows of salt water using the combined potential probe described in [15]. These experiments revealed spontaneous reflectional symmetry breaking (SRSB) in turbulent flows with reflectionally symmetrical initial and boundary conditions. The phenomenon of SRSB is exhibited, in particular, by production of the nonzero mean helicity. Remarkably, experiments showed that spectra of the generated helicity have the maxima in different length scales under the same flow conditions. This variability is in contrast with the behavior of other flow characteristics. The authors of this study believed that generation of the nonzero mean helicity is an intrinsic property at many turbulent flows.

Turbulence energy spectra having a slope $-7/3$ in MHD turbulence were detected first in the experiments conducted by Branover et al.[18]. Examining experimental studies which reported turbulence spectra with the slope close to $-7/3$ in different flows demonstrated that these spectra were not exotic but occurred in various turbulent flows (see the review in Branover et al.[19]). Eidelman et al.[20] reviewed experimental studies of laboratory and atmospheric turbulence, which reported turbulent spectra having the slope close to $-7/3$, and analyzed helical properties of these flows (see also Moiseev and Chkhetiani[21], Koprov et al.[22] and Moiseev et al.[23]). In particular, turbulent velocity spectra with the slopes $-2.3$ and $-2.44$ were detected using velocity structure functions measured in stratified rotating turbulence by Griffiths and Hopfinger[24] (see Fig. 10a in [24]). Turbulence spectra with a slope close to $-7/3$ were obtained in the experiments conducted in stably stratified turbulence by Itsweire and Helland[25] (see Fig. 5 in [25]). We found that wind spectra reported by Nastrom et al.[26] in the analysis of the mesoscale range of Global Atmospheric Sampling Program data also have the slope $-7/3$ (see Fig. 6 in [26]). These steep spectra correspond to velocity fluctuations scales from $15-23$ km down to $2-3$ km.

Asymptotic behavior of the structure function of turbulent velocity field in the inertial and dissipative intervals was studied for the case when the energy and helicity cascades exist simultaneously by Golbraikh and Moiseev[27], Chen[28], Goldbrakh[29], Golbraikh and Eidelman[30], Chkhetiani and Golbraikh[31]. In this case turbulence spectra may have the slopes which are different from $-5/3$ or $-7/3$.

Measurements of helicity were conducted in different types of air and liquid turbulent flows, e.g. in grid turbulence, turbulent wakes and jets, turbulent boundary layer, and turbulent mixing layer. Although these studies revealed some interesting features of helical turbulence they were not continued. One of the possible reasons for this lack of new experiments for direct measurements of helicity is the complexity of the employed probes.

Clearly, further experimental investigations of properties of helical turbulence and of dynamical significance of helicity in turbulence are required.

As we have mentioned above it is feasible to employ another approach for determining helicity of turbulence based on turbulent velocity cross-spectra. The correlation tensor of the velocity $R_{ij}(\mathbf{r},t) = \langle \mathbf{u'}_i(0,t) \cdot \mathbf{u'}_j(\mathbf{r},t) \rangle$ of isotropic turbulence can be represented in the Fourier space as follows:

$$\hat{R}_{ij}(\vec{k}) = A(k)(\delta_{ij} - \frac{k_i k_j}{k^2}) + G(k)\varepsilon_{ijk}\frac{k_k}{k}, \tag{2}$$

where $\delta_{ij}$ the Kronecker tensor, and $\varepsilon_{ijk}$ is the Levi-Civita completely antisymmetric tensor, and gyrotropic last term in the right side is associated with the density of helicity (see e.g. [32]). The mean helicity is given by the following expression:

$$H_t = -\frac{1}{2}\varepsilon_{ijk}\int \hat{R}_{ik}(k)k_j d\vec{k} \tag{3}$$

It is necessary to take into account a range of scales, where turbulence is close to isotropic using cross-spectra of velocity.

There is the lack of experimental investigations of helical turbulence in flows with a rotation and with non-zero mean helicity, and in other flows with coherent structures, where one can expect a pronounced helicity of turbulence. In this study we explore helical turbulence in a Rayleigh-Bernard convection cell with a large scale circulation (LSC). In Section II we describe the experimental set-up and measurement techniques. In Section III we present and discuss the obtained experimental results. Section III also includes analysis of properties of helical turbulence in the observed flow. In Section IV we summarize the obtained results and discuss their relation with the helicity of the flow.

**II. EXPERIMENTAL SET-UP**

The experiments in unstably temperature stratified turbulence were conducted in rectangular cells with dimensions $26 \times 58 \times 26$ cm$^3$ and $26 \times 58 \times 13$ cm$^3$. Hereafter, we use the following system of coordinates: $z$ is the vertical axis and the $y$ axis is directed along the longest wall (see Fig. 1). Aspect ratios of the chambers $A = Y/Z = 2.23$ and 4.46, where $Y$ and $Z$ are the sizes of the cell along $y$ axis and $z$ axis, respectively. The transparent walls of the cell are made of Perspex with the thickness of 10 mm. We present here only basic details of the experimental se-tup that was described comprehensively in [33].

A vertical mean temperature gradient in the turbulent air flow was formed by attaching two aluminum heat exchangers to the bottom (heater) and top (cooler) walls of the test section. The top plate is a bottom wall of the tank with cooling water. Temperature of water circulating through the tank and the chiller is kept constant within 0.1 K. The bottom plate is attached to the electrical heater that provides constant and uniform heating. The temperatures of the conducting plates were measured with four thermocouples attached at the surface of each plate. The temperature difference between the top and bottom plates $\Delta T$ in the experiment was 50 K, and Rayleigh number $Ra = 0.9 \cdot 10^8$ for the cell with the aspect ratio $A = 2.26$.

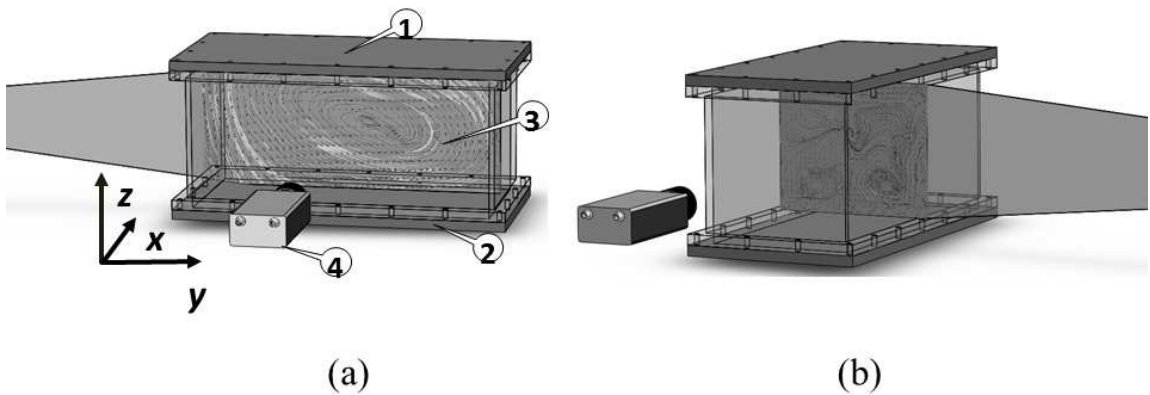

**FIG. 1.** Experimental setup: (1) top heat exchanger (cooler); (2) bottom heat exchanger (heater); (3) walls of the cell; (4) CCD camera. *YZ* (a) and *XZ* (b) cross-sections of athe flow are shown with streamlines.

Velocity fields were measured using Particle Image Velocimetry (PIV) with LaVision Flow Master III system. Double-pulsed light sheet was produced by PIV Nd:YAG laser and light sheet optics with tunable divergence and adjustable focus length. Images were captured with two progressive-scan 12 bit digital CCD cameras with pixel size mcaptured images. Incense smoke of submicron spherical particles with a mean diameter of the order of $0.7\ \mu m$ was used as tracer for the PIV measurements. The software package Lavision DaVis 7 was applied to control all hardware components and for 12 bit image acquisition and visualization.

The velocity field was measured in the two perpendicular cross-sections: in five *YZ* planes and in ten *XZ* planes, so that the distance between neighboring planes was 5 cm (see Fig. 1). Series of 130 and 260 pairs of images acquired with the frequency of 1 Hz were stored for calculating the velocity maps and for ensemble and spatial averaging of turbulence characteristics. The center of the probed flow region coincides with the center of the cell. The probed flow area in the convective cell having the aspect ratio A = 2.23 in *YZ* cross-section was $492 \times 212 \text{ mm}^2$ with a spatial resolution of 302 μm/pixel and in *XZ* cross-section was $237 \times 211 \text{ mm}^2$ with a resolution of 264 μm/pixel. Similarly, in the cell with the aspect ratio $A = 4.46$, the probed flow area is $547 \times 127 \text{ mm}^2$ with the resolution of 278 μm/ pixel in *YZ* cross-section and the probed flow area is $256 \times 125 \text{ mm}^2$ with a resolution of 130 μm/ pixel in *XZ* cross-section.

The obtained PIV images were analyzed with the interrogation window of $32 \times 32$ pixels whereby velocity vector was determined in every interrogation window in the velocity map. The maximum tracer particle displacement in the experiment was about of the quarter of the interrogation window, and the average displacement of tracer particles was about 2.5 pixels. The relative error of the velocity measurements was of the order of 4% for the accuracy of the correlation peak detection with subpixel resolution of 0.1 pixel.

Mean and r.m.s. velocities, spectra, two-point correlation functions, and integral scales of turbulence were determined from the measured velocity fields. The mean and rms velocities for each point of the velocity map (up to 3280 points) were determined by averaging over 130 independent maps and then averaged over 3,280 points when spatial averaging was required. The two-point correlation functions and spectra of the velocity field were determined in the central part of the velocity map by averaging over 130 independent velocity maps. The integral scale of turbulence was determined from the two-point correlation functions of the velocity field. The characteristic turbulence time in the experiments was $\tau_z = 0.8 - 1.2 \text{ s}$, while the characteristic time of the LSC was by one order larger than $\tau_z$. These two characteristic times are much smaller than a time of measurement of velocity or temperature series.

## III. RESULTS AND DISCUSSION

In Fig. 2 we show three different mean flow structures found in the large-scale circulation (LSC). These structures are detected using sets of *YZ* and *XZ* velocity maps of the mean flow similar to those shown in Fig. 3. Inspection of Fig. 2 reveals the main roll having the vorticity aligned in *x* direction and with a length scale of the order of the size of the convection cell. The roll is observed in the velocity map measured in the central *YZ* cross-section of the flow shown in Fig. 3a. This LSC in the flow was observed in our previous experiments in Rayleigh-Bénard convective cell (see Bukai et al. [33]).

However, to the best of our knowledge the existence of two eddy rings adjacent to the bottom and top of the main roll and elongated in *y* direction is reported here for the first time. The eddy rings are showed in the velocity map measured in the central *XZ* cross-section of the flow in Fig. 3b. The length scale of the vortex in these two eddy rings is by a factor of 4÷5 smaller than the vertical length scale of the main roll. The main roll is oriented along one of the diagonals of the cell leaving vacant the corners adjacent to another diagonal of the cell in *YZ* cross-sections of the cell. The eddy rings are formed in these vacant corners and in the adjacent regions of the flow (see Fig. 2).

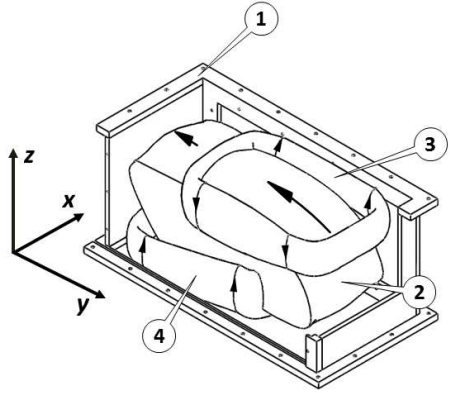

**FIG. 2.** Large scale circulation flow with flow streamlines in Rayleigh-Bernard convective cell with aspect ratio $A = 2.23$: 1– convective cell, 2 – main roll, 3 – top eddy ring, 4 – bottom eddy ring.

Orientation of the main roll depends on the direction of its rotation. If the direction changes, the roll changes orientation along other diagonal, and the top and bottom eddy rings change their positions and occupy other vacant corners.

Similar flow pattern was observed in the case with aspect ratio $A = 4.46$, although in this case the LSC comprises two main rolls having the same vertical velocity direction in a central part of the flow. A schematic view of such two-roll flow pattern with adjacent eddy rings is shown in Fig. 4. There are two eddy rings on the top of the rolls similar to the eddy ring (3) in Fig. 2 which occupy top vacant corners, and one eddy ring at the bottom formed by merger of two eddy rings similar to the eddy ring (4) in Fig. 2. If the direction of rotation of the main rolls is reversed, a position of the rings also changes to the opposite whereby two eddy rings are located at the bottom and one eddy ring is located at the top.

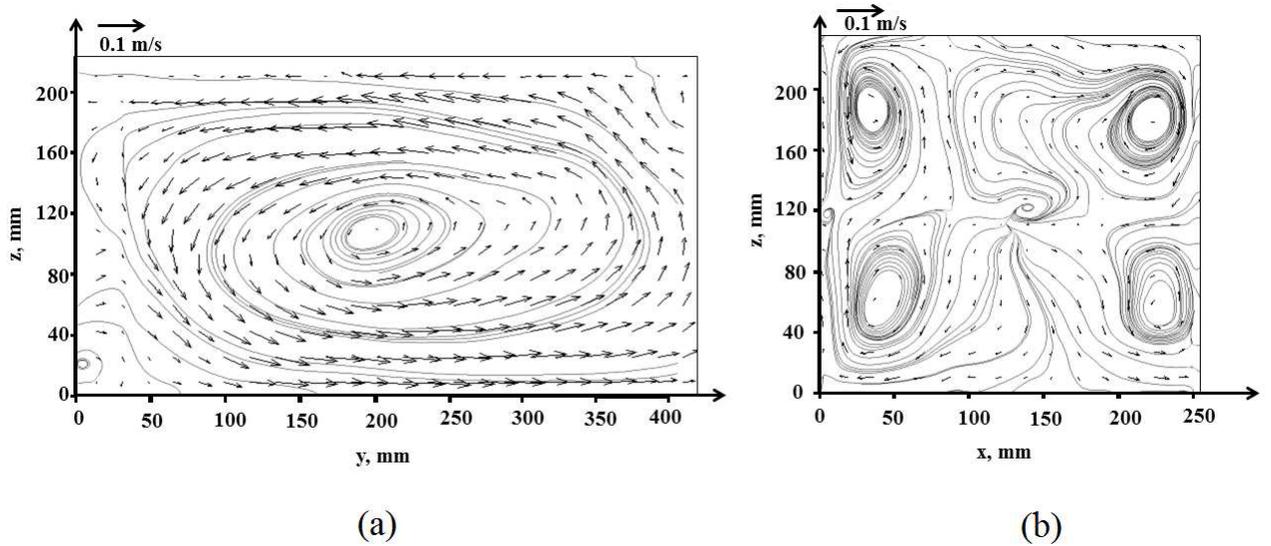

(a)                      (b)

**FIG. 3**. Velocity vector maps and streamlines of the mean flow: (a) in the central *YZ* cross-section, (b) in the central *XZ* cross-section.

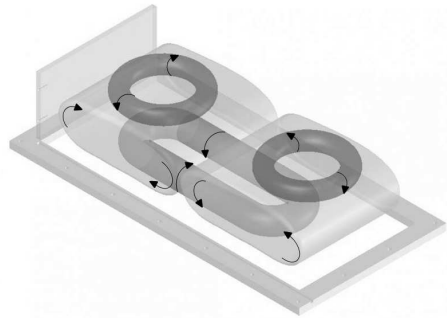

**FIG. 4.** Scheme of large-scale circulation in cell with aspect ratio $A = 4.46$.

Similar asymmetry of the main roll in the LSC was reported in numerous experiments conducted in convective cells after the study by Qiu and Tong [34] who detected asymmetric orientation of the LSC in cylindrical convective cells with aspects ratios $A = 1$ and $A = 0.5$. This asymmetry of the LSC was observed in cylindrical and rectangular convective cells having different aspect ratios $A = 0.5 - 4.46$, including experiments reported in the present study. Since our experiments also revealed the existence of the additional eddy rings in the LSC, it is plausible to suggest that the main roll and adjacent eddy rings are intrinsic feature of the LSC in temperature stratified convective flows.

In the following we analyze in more details a flow pattern in a convective cell with aspect ratio $A = 2.23$. The vertical distributions of the mean flow velocity $U_y$ measured in the intersections of five $YZ$ velocity maps with a central $XZ$ cross-section, are showed in Fig. 5. Analysis of the horizontal velocity $U_y$ distributions in the $YZ$ cross-sections with different coordinates $x$ shows that they are almost identical. The vertical distributions of $U_y$ attains the extrema at the distance of approximately $\Delta z/Z = 0.1$ from the top and bottom walls. Horizontal distributions of the mean vertical velocity $U_z$ in the central $YZ$ cross-section of the flow at different values of $z$ are similar. The vertical velocity is small in the central part of the flow and increases at the distances $\Delta y/Y < 0.15$ from the side walls.

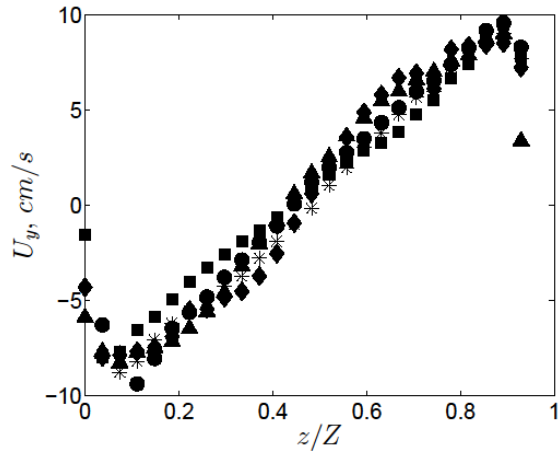

**FIG. 5**. Mean velocity $U_y$ distributions in intersections of five $YZ$ maps with a central $XZ$ cross-section at coordinates $x/X = 0.88$ (triangles), 0.65 (squares), 0.45 (circles), 0.26 (stars) and 0.065 (diamonds).

Vertical distributions of vorticity of the mean flow $\omega_y$ measured in the central $XZ$ plane of the velocity map (see Fig. 3b) are shown in Fig.6. The distributions were measured near the $XZ$ plane intersections with two $YZ$ planes located at the distance $\Delta x/X \sim 0.08$ from the frontal and back walls of the convective cell. Inspection of Fig. 6 reveals that vorticity of the top and bottom eddy rings attains extrema at the distance of about $\Delta z/Z = 0.25$ from the top and bottom walls.

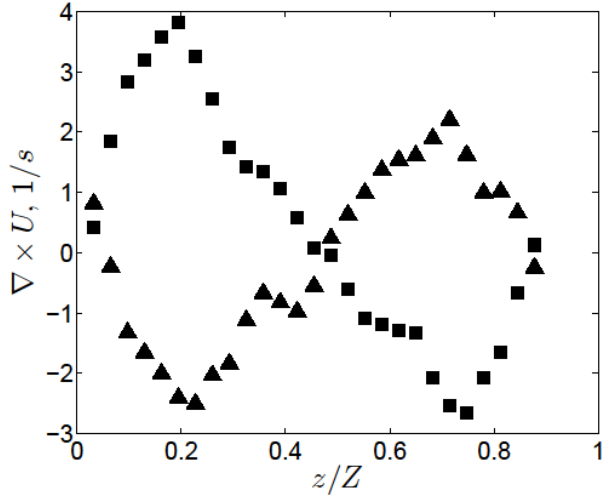

**FIG. 6.** Distributions of vorticity of mean velocity in the central *XZ* cross-section measured at $x/X = 0.065$ (squares) and $0.845$ (triangles).

The mean horizontal velocity of the main roll $U_y$ and the mean vorticity of two eddy rings $\omega_y$ are aligned in the significant part of the flow (see Fig. 1) similar to the Beltrami flow (see Wang [35]). Our estimations revealed that the contribution of $H_{my} = U_y \cdot \omega_y$ to the helicity of the mean flow $H_m$ is significantly larger than other contributions, $H_{mx}$ and $H_{mz}$. Two components of velocity, $U_z$ and $U_x$, are by a factor 3 smaller than $U_y$. Since the vorticity component $\omega_x$ is by a factor of 3 smaller than $\omega_y$, and $\omega_z$ is even smaller than $\omega_x$, the term $H_{my}$ is a fairly good estimate of the helicity of the mean flow $H_m$. Hereafter $H_m$ is determined using five *YZ* maps of the horizontal mean velocity $U_y$ and ten *XZ* maps of vorticity component $\omega_y$ of the mean velocity fields. Polynomial fit is used to determine distributions of the mean velocity component $U_y$ with the same resolution and for the same locations as $\omega_y$. Averaged over all *XZ* maps distribution of the helicity of the mean velocity $H_m$ attains the extrema in four quadrants of the *XZ* cross-section whereby the sign of helicity alternates and is the same for the diagonal quadrants. The main contribution to helicity (about 75%) comes from the external quarter of each quadrant.

Horizontal ($z/Z = 0.2$) and vertical ($x/X = 0.065$) distributions of $H_m$ averaged over *XZ* cross-sections are shown in Fig. 7. Inspection of the distribution of the mean flow helicity $H_m$ in a *XZ* cross-section shows that the signs of the mean helicity in the left ($H_m < 0$) and in right ($H_m > 0$) parts of the flow are opposite.

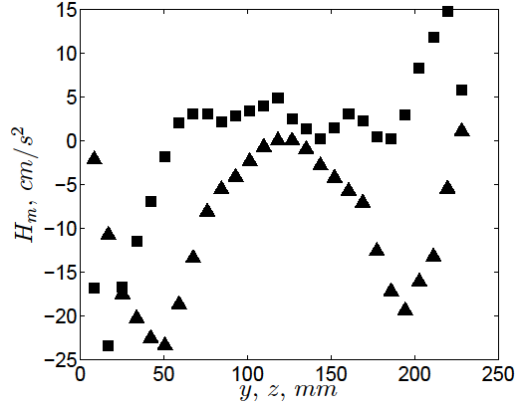

**FIG. 7.** Horizontal (squares) and vertical (triangles) distributions of helicity of the mean flow $H_m$ in the central *XZ* cross-section.

Spatial spectra of vertical distributions of the mean velocity component $U_y$ and helicity of the mean flow are shown in Fig. 8. Inspection of this figure reveals that the mean velocity spectrum has a pronounced maximum at the scale of about 25 cm which is close to the size of the *XZ* cross-section of the cell. Steep decrease of the spectrum for larger and smaller scales indicates a local forcing of turbulent energy.

Spectrum of the helicity of the mean flow $H_m$ is also showed in Fig. 8 where a spectral density is approximately constant in the range of large scales from 25 cm to 8 cm, and then sharply decreases at smaller and larger scales. It can be hypothesized that helicity forcing of turbulence occurs in a wide range of scales in contrast to the localized spectrum of the mean velocity $U_y$. Local forcing of energy and non-local forcing of helicity affects turbulence properties as discussed below.

We estimated homogeneity of turbulent energy distribution in the flow. Distributions of the normalized turbulent kinetic energy are shown in Fig. 9. They are averaged in the horizontal or vertical directions in the central *YZ* and *XZ* cross-sections and normalized by the mean value of turbulent kinetic energy, $u_m^2$. The ratio of spatially averaged turbulent energies in the *YZ* and *XZ* cross-sections is 1.04 while r.m.s of the spatial fluctuations of the turbulent kinetic energy in YZ cross-section $\sigma = 18.5\,\%$ and $\sigma = 14\,\%$ in the *XZ* cross-section. Consequently, turbulent energy distribution is close to homogeneous and isotropic although a weak dependence on the LSC is observed.

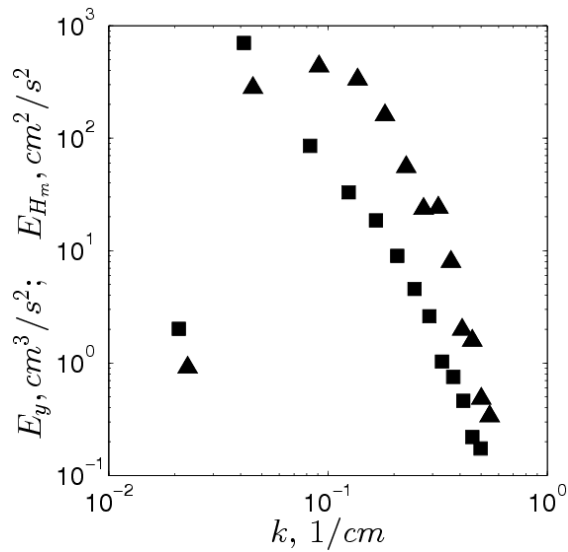

**FIG. 8.** Spectra of vertical distribution of the horizontal mean velocity component, $U_y$, in the central *YZ* cross-section, $E_y$ (squares) and of the helicity of the mean flow, $H_m$ in the central *XZ* cross-section, $E_{Hm}$ (triangles).

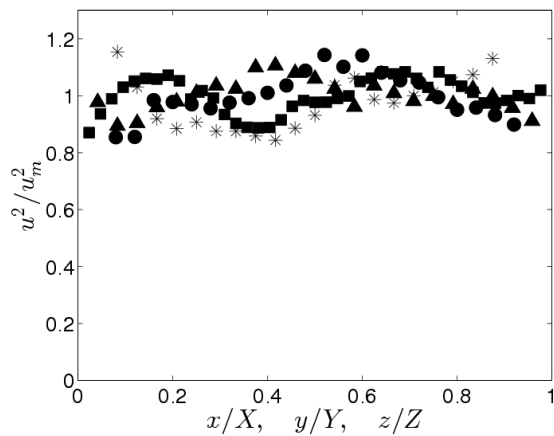

**FIG. 9.** Normalized by the mean value turbulent kinetic energy averaged over vertical (squares) and horizontal (triangles) directions in the *YZ* cross-section and over vertical (stars) and horizontal (circles) directions in the *XZ* central cross-sections.

Energy spectra of components of turbulent velocity (for $u'_y$ in *YZ* and $u'_x, u'_z$ in *XZ* central cross-sections) are shown in Fig. 10. Inspection of Fig. 10 reveals the existence of two sub-ranges in the spectra with the exponents close to the Kolmogorov's $-5/3$ for larger scales and $-7/3$ for smaller scales. Transition between these sub-ranges in the spectra occurs at the characteristic length scale $L_S$ of the order of $6.5 - 7.5$ cm. The inertial range of energy spectra with the slope $-5/3$ begins at the scale of ~15 cm and extends down to $L_S$ due a local energy forcing by the mean flow at the scale of 25 cm (see Fig. 8). In this spectral range the slope of the energy spectra is not affected by helicity forcing in scales $8 - 25$ cm (see Fig. 8).

The adjacent inertial sub-range with a slope of $-7/3$ begins at the scale $L_S$, where forcing of helicity of turbulence by the mean flow helicity almost vanishes, and extends down to the scales of the order of ~ 2.5 cm. It is conceivable that $-7/3$ spectrum extends to smaller scales although we have not succeeded to determine the slope in this range because of aliasing errors.

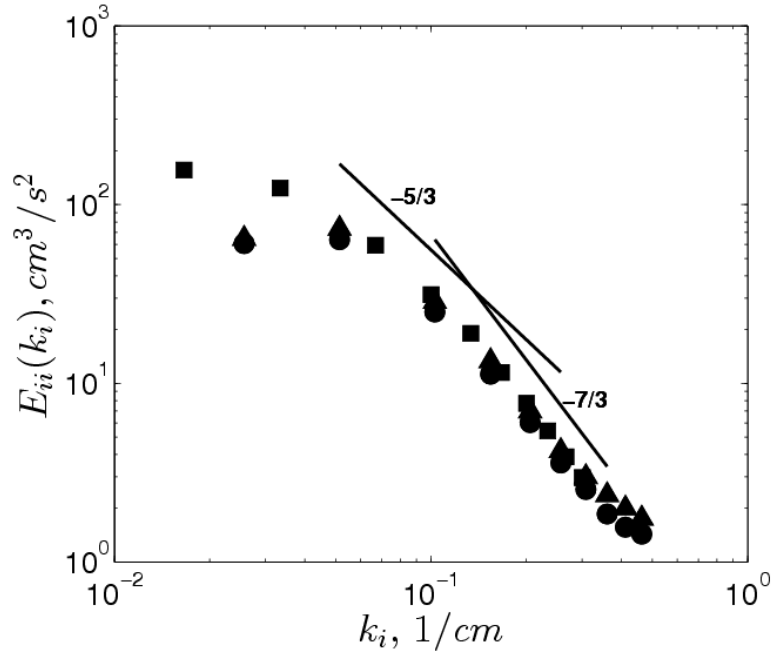

FIG.10. Energy spectra of *x* (triangles), *y* (squares), and *z* (circles) components of turbulent velocity measured in the *YZ* and in the *XZ* central cross-sections in the flow.

Inspection of turbulence spectra in Fig. 10 shows that turbulence at the scales smaller than 20 cm is close to isotropic. Anisotropy of turbulence at larger scales in the *YZ* cross-section is associated with the large-scale flow in the cell having the aspect ratio $A = 2.23$. Nonzero mean helicity of turbulence affects energy spectrum at the scales larger than the scale determined by the ratio of turbulent energy to helicity of turbulence (Belian et al.[10]). Notably, this scale in our experiments was of the order of 20 cm.

We have also verified the isotropy of the flow using the known relationship between the longitudinal and transverse components of the velocity correlation tensor, $R_{xx}(x)$ and $R_{yy}(x)$:

$$R_{yy}(x) = \frac{1}{2x} \frac{d}{dx}\left(x^2 R_{xx}(x)\right) \tag{4}$$

It is more convenient to examine the longitudinal and transverse components of the velocity correlation tensor in the Fourier space. Equation (4) in Fourier space reads (see e.g. Monin & Yaglom [36], Section 12.3):

$$E_{yy}(k_x) = \frac{1}{2}\left(E_{xx}(k_x)\right) - k_x \frac{dE_{xx}(k_x)}{dk_x}, \tag{5}$$

where $E_{xx}(k_x)$ and $E_{yy}(k_x)$ are spectral energy densities of turbulent velocity components. Inspection of Fig. 11 shows that at spatial scales less than 15 cm calculated and measured transverse velocity spectra are close, i.e. turbulence is locally isotropic.

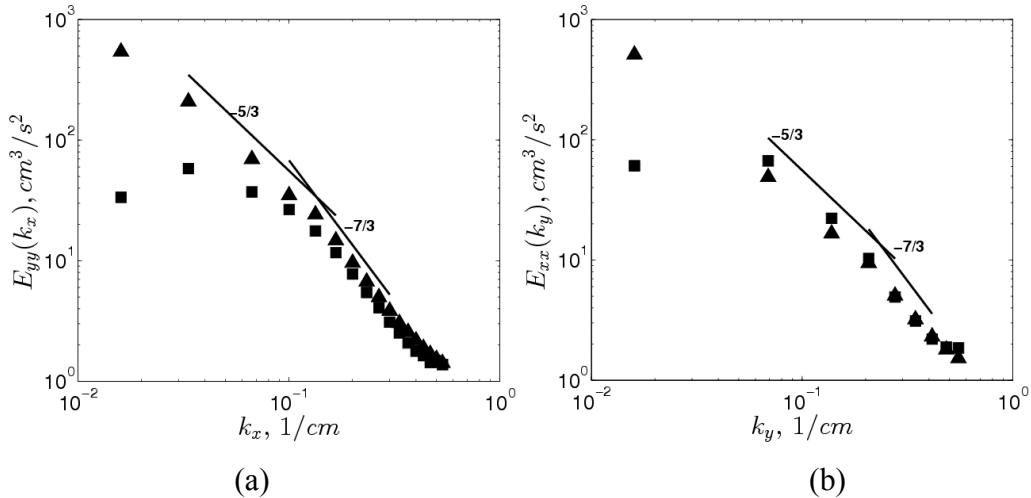

(a)           (b)

**Fig.11**. Comparison of the transverse spectra, $E_{yy}(k_x)$ (a) and $E_{xx}(k_y)$ (b). Squares denote measured spectra. Triangles denote calculated spectra (see Eq. (5)) using the measured longitudinal spectra $E_{xx}(k_x)$ and $E_{yy}(k_x)$.

Cross-spectra of turbulent velocity (see Fig.12a) and spectra of turbulent helicity (see Fig.12b) were determined in *YZ* and *XZ* cross-sections. Transition between two different slopes of the cross-spectra occurs at the scales of the order of 6 – 7 cm that are close to $L_S$. For the scales larger than $L_S$ where energy spectra slope is close to $-5/3$, the slope of helicity spectra is close to -2/3. In the range of scales smaller than $L_S$, cross-spectra and spectra of velocity have the slopes $-7/3$ while the slope of helicity spectra is $-4/3$. These spectra slopes are pertinent to the flows where the turbulent helicity determines the behavior of the spectra of energy and helicity (see Moiseev & Chkhetiani [21], Chkhetiafni & Golbraikh [31]). The magnitude of the turbulent helicity is estimated by integrating helicity spectra in the range of length scales of isotropic turbulence as 0.45 – 0.5 cm/s$^2$. Notably, the length scales 6.8 – 8.3 cm obtained here as the ratio of turbulent energy to turbulent helicity are close to $L_S$.

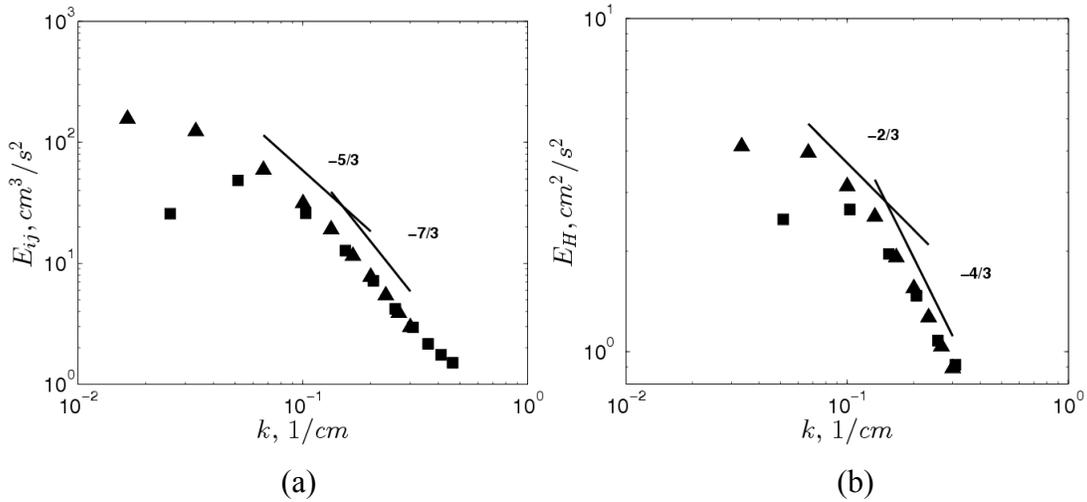

**Fig. 12.** Cross spectra of turbulent velocity (a) and spectra of turbulent helicity (b) in the central *YZ* (triangles) and *XZ* (squares) cross-sections.

It must be noted that helicity spectra having the slope $-2/3$ were discussed previously by Brissaud et al. [11] using assumption that energy and helicity were injected into turbulence cascade in the opposite directions. Turbulent energy is transported from the injection scale to larger scales and helicity is transferred from injection scale to smaller scales. Similar helicity spectra were discussed by Lesieur [32], Golbraikh [28] without invoking this assumption. In these studies helicity did not affect energy cascade expressed in terms of the energy dissipation rate what implied that the Kolmogorov's energy cascade was not modified by the helicity.

Some numerical studies of helicity (see e.g. Sanada [37], Chen et al. [27], Stepanov et al. [38] and references therein) and experiments conducted by Koprov et al. [22] showed that in the range of scales where the energy spectra slopes are close to $-5/3$, the helicity spectra have the same slope. It must be noted that commonly in numerical simulations the turbulence and helicity forcings are localized at the same spatial scale. However, spatial distribution of turbulence and helicity forcing in the experiments reported in this study are quite different. Indeed, the turbulent energy source is localized in the scale close to 25 cm while the source of helicity has maximum in the scale close to 11 cm and extends over the range of scales from 8 to 25 cm (see Fig. 8). The inertial range of energy spectra, having the slope close to $-5/3$ in this range, is determined by the turbulent energy cascade.

Inertial ranges of energy and helicity spectra having the slopes close to $-7/3$ and $-4/3$, respectively, for scales smaller than $L_s$ are determined by the helicity cascade (see Fig. 10). Notably, similar results have been obtained in numerical simulations which used shell model with a constant flux of energy and non-local helicity forcing (Shestakov et al. [39]). These simulations yielded spectra of energy and helicity having the slopes $-7/3$ and $-4/3$, respectively.

**IV. CONCLUSIONS**

We studied turbulence in two rectangular Rayleigh-Bénard convective cells having different aspect ratios. In the experiments we observed large-scale circulation (LSC) that includes one or two main rolls for the cell aspect ratios 2.23 and 4.46, respectively, and elongated eddy rings adjacent to the bottom and top of the cell. Similarly to the large-scale flow in the cylindrical convective cell, the observed rolls are orientated in the direction of one of the diagonals of the cell, and the eddy rings occupy the vacant corners adjacent to other diagonal. Notably, when direction of rotation of the LSC changes the rolls also change their orientation to the other diagonal of the cell. We hypothesize that the appearance of eddy rings in Rayleigh-Bénard convective cells is associated with the asymmetric orientation (along one of the main diagonals) of the main roll in the LSC, and is an intrinsic feature of such flows. We found that helicity of the observed mean flow is relatively large. The latter observation is in compliance with the topology of the LSC which is similar to the Beltrami flow in some regions in the flow.

Spatial distribution of the turbulent kinetic energy density is almost locally isotropic in the central region of the cell. The measured velocity spectra and cross-spectra in the inertial scale range exhibit two different power law sub-ranges having the exponents close to $-5/3$ and $-7/3$. The helicity of turbulence and its spectra were determined using cross-spectra of the measured turbulent velocity in the cell. The slopes of the power law spectra of helicity in the intervals above and below the transition length scale $L_S$ are close to $-2/3$ and $-4/3$, respectively. The transition between these two sub-ranges in the measured spectra and cross spectra occurs at the length scale of $L_S = 0.27Z$ which is approximately equal to the ratio of the turbulent kinetic energy and helicity of isotropic turbulence. We believe that emergence of these two power law sub-ranges is associated with energy and helicity cascades that affect turbulence.

We analyzed interplay between the mean flow and turbulence using spatial spectra of the mean velocity and mean flow helicity, turbulent energy and turbulent helicity spectra. There is a pronounced maximum in the spectrum of the mean velocity at the scale of about 25 cm that is close to the size of the cell, and steep decrease in larger and smaller scales. This behavior is a clear indication of the local turbulent energy forcing. The measured mean flow helicity spectra are almost constant for the length scales between 8 cm and 25 cm and sharply decrease outside this range. Consequently, it can be hypothesized that the mean flow helicity is a source of turbulent helicity forcing in this range. The injection of helicity in the range of length scales larger than $L_S$ does not affect the slope of the energy spectra which is close to the Kolmogorov's $-5/3$. It can be argued that the presence of turbulent helicity causes the change of the energy spectra slope to $-7/3$ in the scales smaller than $L_S$ where spectral density of energy decreases.

The flow in the Rayleigh-Bénard convective cell is an example of a flow where nonzero helicity of the mean and turbulent flows determine the properties of turbulence, in particular, the slopes of the energy and helicity spectra. Remarkably, similar inertial sub-ranges in the turbulent energy spectra were observed in geophysical turbulent flows and in rotating turbulence. Therefore it is conceivable to suggest that helicity can be observed also in other turbulent flows with coherent structures.

**Acknowledgements.** This work has been supported by the Israel Science Foundation governed by the Israeli Academy of Sciences (Grant No. 1037/11).